\documentclass[prb,preprint, superscriptaddress, nolongbibliography]{revtex4-2}

\usepackage[dvipdfmx]{graphicx}
\usepackage{dcolumn}
\usepackage{bm}
\usepackage{color}

\begin{document}

\title{Two-dimensional heavy fermion in a monoatomic-layer Kondo lattice YbCu$_2$}
\author{Takuto Nakamura}
\email{nakamura.takuto.fbs@osaka-u.ac.jp}
\affiliation{Graduate School of Frontier Biosciences, Osaka University, Suita 565-0871, Japan}
\affiliation{Department of Physics, Graduate School of Science, Osaka University, Toyonaka 560-0043, Japan}
\author{Hiroki Sugihara}
\affiliation{Department of Physics, Graduate School of Science, Osaka University, Toyonaka 560-0043, Japan}
\author{Yitong Chen}
\affiliation{Department of Physics, Graduate School of Science, Osaka University, Toyonaka 560-0043, Japan}
\author{Ryu Yukawa}
\affiliation{Graduate School of Engineering, Osaka University, Suita 565-0871, Japan}
\author{Yoshiyuki Ohtsubo}
\affiliation{National Institutes for Quantum Science and Technology, Sendai 980-8579, Japan}
\author{Kiyohisa Tanaka}
\affiliation{Institute for Molecular Science, Okazaki 444-8585, Japan}

\author{Miho Kitamura}
\affiliation{Photon Factory, Institute of Materials Structure Science, High Energy Accelerator Research Organization (KEK), 1-1 Oho, Tsukuba 305-0801, Japan}
\author{Hiroshi Kumigashira}
\affiliation{Institute of Multidisciplinary Research for Advanced Materials (IMRAM), Tohoku University, Sendai, 980–8577, Japan}
\author{Shin-ichi Kimura}
\email{kimura.shin-ichi.fbs@osaka-u.ac.jp}
\affiliation{Graduate School of Frontier Biosciences, Osaka University, Suita 565-0871, Japan}
\affiliation{Department of Physics, Graduate School of Science, Osaka University, Toyonaka 560-0043, Japan}
\affiliation{Institute for Molecular Science, Okazaki 444-8585, Japan}

\date{\today}

\maketitle
\textbf{
The Kondo effect between localized $f$-electrons and conductive carriers leads to exotic physical phenomena.
Among them, heavy-fermion (HF) systems, in which massive effective carriers appear due to the Kondo effect, have fascinated many researchers.
Dimensionality is also an important characteristic of the HF system, especially because it is strongly related to quantum criticality [S. Sachdev, Science 288, 475 (2000)].
However, perfect two-dimensional (2D) HF materials have not been reported yet.
Here, we report the surface electronic structure of the monoatomic-layer Kondo lattice YbCu$_2$ on a Cu(111) surface observed by synchrotron-based angle-resolved photoelectron spectroscopy.
The 2D conducting band and the Yb 4$f$ state, located very close to the Fermi level, are observed.
These bands are hybridized at low-temperature, forming the 2D HF state, with an evaluated coherent temperature of about 30 K.
The effective mass of the 2D state is enhanced by a factor of 100 by the development of the HF state.
Furthermore, clear evidence of the hybridization gap formation in the temperature dependence of the Kondo-resonance peak has been observed below the coherent temperature.
Our study provides a new candidate as an ideal 2D HF material for understanding the Kondo effect at low dimensions.
}

Heavy fermion (HF) systems in rare-earth (RE) intermetallic compounds originating from hybridization between localized $f$-electrons and conduction electrons, namely $c$-$f$ hybridization, are central topics in the field of the strongly-correlated electron systems \cite{HFcolman}.
At low temperatures, depending on the strength of the $c$-$f$ hybridization, the physical properties change from itinerant $f$ electrons because of the Kondo effect or magnetic order originating from the magnetic moment of localized $f$ electrons due to Ruderman-Kittel-Kasuya-Yosida (RKKY) interactions.
The competition between itinerant and localized characters of the $f$-electrons make a quantum critical point (QCP), resulting in the emergence of fertile quantum phenomena such as non-Fermi liquid behavior, and non-BCS HF superconductivity \cite{SCf, HFSC}.

On the other hand, the dimensionality in the system characterizes the fundamental physical property.
In low-dimensional systems, the enhancement of the electron-electron correlation and/or breaking of the inversion symmetry leads to novel quantum states such as Rashba-type spin-splitting \cite{Rashba}, Tomonaga--Luttinger liquid \cite{Tomonaga, Luttinger}, and unconventional superconductivity \cite{monoNbSe2SC, IsingSC}.
The combination of the HF state and low dimensionality modifies the ground state of the system because the order parameter of these systems is much more sensitive to dimensionality \cite{HFdimension}.
The ground state of two-dimensional (2D) HF can be easily controlled to the vicinity of a quantum critical point, which is the host to realize unconventional physical properties such as HF superconductivity, by simple external fields such as gate-tuning \cite{MoireKondo, CeCoIngate}, and surface doping \cite{2DCeSiI} in addition to traditional external perturbations; temperature, pressure, and magnetic field.
The fabrication of artificial low-dimensional strongly-correlated electron systems is one of the suitable methods to investigate the novel electronic phase.
In the Ce-based artificial superlattice, the suppression of antiferromagnetic (AFM) ordering as well as the increase of the effective electron mass with decreasing of the thickness of the Ce-layer \cite{CeIn3-2D} and the emergence of the strong-coupling superconductivity \cite{Ce2DSc} have been reported.
To understand the fundamental properties of 2D HF systems, it is necessary to clarify the electronic band structure and the formation mechanism of the HF.
However, the details have remained unclear due to the lack of promising materials and the extremely low transition temperatures of less than a few K to HF even in known materials \cite{CeIn3-2D, CeRhInYbRhIn, He3}.

The growth of the well-ordered atomically thin film on a single crystal substrate is a suitable technique to access such 2D electron systems.
So far, various 2D Kondo-lattice has been fabricated on the substrates; multilayer CePt$_5$ thin-film on Pt(111) \cite{CePt5ARPES},  CePb$_3$ on Si(111) \cite{CePb3}, Graphene on SmB$_6$ \cite{GrapheneSmB6}, and a checkerboard pattern of organic molecules on Au(111) \cite{OMKondo}.
However, Yb-based 2D HF material, in which the Yb ion is the most fundamental element to realize HF \cite{YbAlB4ARPES, YbRh2Si2ARPES} and has a symmetrical electronic-hole configuration to the Ce one, has not been reported.
In particular, the RE-based monoatomic layer Kondo-lattice showing HF state has never been reported.

In this study, we report the HF electronic structure of a novel Yb-based monoatomic layer Kondo lattice; synchrotron-based angle-resolved photoelectron spectroscopy (ARPES) on monoatomic layered YbCu$_2$ on Cu(111).
The surface atomic structure of the YbCu$_2$ on Cu(111) is shown in Fig. 1(a).
The Yb atoms surrounded by Cu atoms are arrayed in a triangular lattice.
In a similar surface alloy RE NM$_2$/NM(111) (NM = noble metal), various physical properties appear such as FM ordering \cite{GdAu2, GdAu22, EuAu2} and Weyl nodal-line fermion \cite{GaAgnodal} depending on the containing RE element, but there is no report on the appearance of HF character so far.
Figures 1(c) and 1(d) show the LEED patterns of the Cu(111) substrate and the Yb-adsorbed Cu(111) surface at 70 K, respectively.
In addition to the primitive (1$\times1$) spots originating from Cu(111) substrate indicated by yellow allows, the ($\sqrt{3}\times\sqrt{3}$)R30$^{\circ}$ structure with the Moir\'{e} patterns, originating from the small lattice mismatch between Cu(111) and the topmost surface alloy layer, was observed, indicating the successful fabrication of the monoatomic YbCu$_2$ layer, one possible model of the Yb-Cu surface alloy system, on the Cu(111) substrate.
Note that the overall trend of the LEED patterns is consistent with those of other RE NM$_2$/NM(111) systems \cite{GdAg, YbAu2, DyNM, EuAu2, GdAu2, GdAu22}.

The itinerant or localized character of Yb 4$f$ electrons is strongly reflected in the valency of the Yb ions.
Figure 2(a) shows Yb 3$d$ core-level spectrum of YbCu$_2$/Cu(111) at 15 K.
The photoelectron peaks at the binding energies of 1528 and 1538~eV originate from the Yb$^{2+}$ and Yb$^{3+}$ $3d$ final states, respectively, after photoexcitation.
From the intensity ratio between the Yb$^{2+}$ and Yb$^{3+}$ peaks after subtracting the background indicated by the dotted line in the figure, the mean valence of Yb ions was evaluated as 2.41 $\pm$ 0.01.
To confirm the consistency of the coexistence of Yb$^{2+}$ and Yb$^{3+}$ observed in the Yb 3$d$ core-level spectra to the electronic state near the  Fermi level ($E_{\rm F}$), angle-integrated valence-band photoelectron spectra of the Cu(111) clean substrate and the YbCu$_2$/Cu(111) surface are shown in Fig. 2(b).
The Cu 3$d$ states at the binding energy of 3 eV are dominant in the Cu(111) substrate.
In the YbCu$_2$/Cu(111) spectrum, there are two narrow peaks originating from the Yb$^{2+}~4f$ spin-orbit pair near $E_{\rm F}$ and broad peaks of Yb$^{3+}~4f$ final states and Cu~3$d$ states at the binding energy of 3--13~eV.
These results strongly suggest that the Yb ions in monoatomic layer YbCu$_2$ are mixed valence.
Note that in the YbAu$_2$/Au(111), which has a similar atomic structure to YbCu$_2$/Cu(111), Yb ions are almost divalent \cite{YbAu2}.
The reason for the difference in the Yb valence between YbCu$_2$ and YbAu$_2$ would be due to the in-plane lattice compression, which can be explained by the analogy from the bulk Yb-based intermetallic compounds under high pressure \cite{YbNiGapressure}, because the lattice constant of bulk Cu is about 10 \% smaller than that of Au.
The lattice compression would promote the valence transition from Yb$^{2+}$ to Yb$^{3+}$ due to the smaller ionic radius of Yb$^{3+}$ than that of Yb$^{2+}$, realizing a mixed-valence state in YbCu$_2$/Cu(111).

Figures 2(d,e) show ARPES band dispersions at 10 K along $\bar{\Gamma}$--$\bar{\rm K}$ and $\bar{\Gamma}$--$\bar{\rm M}$, respectively, in the hexagonal surface Brillouin zone (SBZ) shown in Fig.~2(c).
The flat band is close to $E_{\rm F}$ and highly dispersive bands are observed near the $\bar{\Gamma}$ point.
According to the previous study and the DFT calculation for other RE NM$_2$/NM(111) families, the flat band and well-dispersive bands mainly originate from the Yb$^{2+}$ 4$f_{7/2}$ and the mixing of the Yb 5$d$ and Cu $sp$ and $d$ orbitals, respectively \cite{GdAu22, YbAu2, ErCu2}.
The detailed assignments of these bands are shown in Supplementary Note 1.
It should be noted that the photoelectron intensities of the dispersive band near the $\bar{\Gamma}$ point at the positive wavenumber region are relatively weak in both  $\bar{\Gamma}$--$\bar{\rm K}$ and $\bar{\Gamma}$--$\bar{\rm M}$ directions due to a photoexcitation selection rule \cite{PEStext}.
The energy position of the Yb$^{2+}$ 4$f_{7/2}$ is very close to $E_{\rm F}$, which is the general feature of the Yb-based HF system such as $\beta$-YbAlB$_4$ \cite{YbAlB4ARPES} and YbRh$_2$Si$_2$ \cite{YbRh2Si2ARPES}, suggesting the same mixed valent character of Yb ions in YbCu$_2$/Cu(111) as the result of Yb 3$d$ peaks.
It should be noted that the surface electronic structure of almost all HF materials tends to be localized, which is inconsistent with that of the bulk, due to a surface lattice expansion.
It should also be noted that the Yb-ions in bulk YbCu$_2$ are mixed-valent \cite{bulkYbCu2PES, BulkYbCu2}, which is similar to YbCu$_2$/Cu(111), but the orthorhombic crystal structure is different from that of YbCu$_2$/Cu(111).
Additionally, the surface state of bulk YbCu$_2$ is divalent, which is consistent with other Yb compounds. Therefore, the origin of the mixed-valent character of YbCu$_2$/Cu(111) is not the same as that of bulk YbCu$_2$.

In Figures 2(d,e), the Yb$^{2+}$ 4$f$ flat band is modulated at the cross points to the conduction bands just below $E_{\rm F}$, which is evidence of $c$-$f$ hybridization \cite{CeCoIn}.
It should be noted that the $c$-$f$ hybridization bands can appear in periodically located Yb and Cu atoms on the surface, not in randomly diluted Yb impurities in bulk Cu.
The hole bands at the $\bar{\Gamma}$ point and the Yb$^{2+}$ 4$f_{7/2}$ states near $E_{\rm F}$ can be confirmed to originate from the YbCu$_2$ layer by calculations \cite{GdAu22, YbAu2, ErCu2} (see Supplimentary Fig.S2).
Figure 2(f) shows a series of the constant energy contours of the YbCu$_2$/Cu(111) surface.
In the map at the binding energy of 0 eV, which corresponds to the experimental Fermi surfaces, there are strong photoelectron intensity areas at $\bar{\Gamma}$ point and near the zone boundary.
However, as shown in Figs. 2(d,e), non-bonding 4$f$ character only appears near the $\bar{\rm K}$ and $\bar{\rm M}$ points, but the $c$-$f$ hybridization feature only exhibits near the $\bar{\Gamma}$ point.
Therefore, in the following part, we focus on the hybridized band around the $\bar{\Gamma}$ point to investigate the detail of the HF character appearing in monoatomically layered YbCu$_2$.

In the HF system, the size of the Fermi surface is modulated by the changing of the temperature due to the enhancement of the $c$-$f$ hybridization.
Figure 3(a) shows the temperature-dependent ARPES images along $\bar{\Gamma}$--$\bar{\rm K}$.
The overall trend of those images is consistent except near $E_{\rm F}$.
To reveal the change of the dispersion at $E_{\rm F}$ in more detail, the momentum-distribution curves (MDCs) at $E_{\rm F}$ are plotted in Fig. 3(b).
The peak position of the hybridized band near the $\bar{\Gamma}$ point at 130 K is shifted toward a higher wavenumber at 7 K, suggesting an enlargement of the Fermi surface by the development of the $c$-$f$ hybridization at low temperature.

Figure 3(c) shows the ARPES image around the $\bar{\Gamma}$ point taken with circularly polarized photons at 15 K.
The $c$-$f$ hybridization branches S1 and S2 are visible.
To determine the dimensionality of the  $c$-$f$ hybridization bands, the photon-energy dependence of ARPES was measured as shown in Fig. 3(d).
Both S1 and S2 bands show no photon-energy dependence, indicating no out-of-plane ($k_z$) dispersion.
These experimental results strongly suggest that the $c$-$f$ hybridization band is formed in the 2D YbCu$_2$ plane.
To evaluate the $c$-$f$ hybridization feature, comparing it to the periodic Anderson model (PAM) is useful \cite{PESPAM}.
In the case of the electron correlation $U_{ff}$ is zero or infinity, the band dispersions ${E_k}^{\pm}$ of PAM  is explained as
\begin{equation}
{E_k}^{\pm} = \frac{\epsilon_c + \epsilon_f  \pm \sqrt{{(\epsilon_c -\epsilon_f)}^2 - 4{V_k}^2}}{2}
\end{equation}
where $\epsilon_c$ and $\epsilon_f$ are the dispersions of the conduction band and the 4$f$ band, respectively, and $V_k$ is the hybridization intensity.
For the fitting by the PAM, a $k$-linear hole band dispersion was assumed for the conduction band to reproduce the steep band shape observed by ARPES.
The fitting results are shown in Fig. 2(c).
The filled and break lines indicate the band dispersions ${E_k}^{\pm}$ with  $V_k$ = 120 meV and 0 meV, respectively.
The detailed process of fitting by the PAM is shown in Supplementary Note 5.
By comparing the experimentally obtained heavy conduction band at 15 K and the simulated bare conduction band, a mass enhancement factor between the effective mass $m^{*}$ of the HF state at 15 K and unhybridized one $m_{b}$ ($m^{*}/m_{b}$) is evaluated as about 120 suggesting the appearance of heavy quasiparticles at low temperatures.

We now discuss the temperature dependence of the quasiparticle peak just below $E_{\rm F}$, so-called Kondo resonance (KR) peak.
In the HF system, the temperature dependences of the energy position and intensity of the KR peak are reflected in the spectral weight transfer between 4$f$ state and the conduction band as well as renormalization due to the development of the $c$-$f$ hybridization.
Figure 4(a1) shows the angle-integrated photoelectron intensity near $E_{\rm F}$ as a function of temperature, and Fig. 4(a2) is the same, but the intensities are divided by the Fermi-Dirac distribution function convolved with the instrumental resolution.
The KR peak energy is shifted to the $E_{\rm F}$ side with decreasing temperature, indicating the evolution of the renormalization due to the HF formation.
To discuss the temperature-dependent development of the HF state in more detail, the peak positions and intensities are obtained from the fitting of Fig. 4(a2) by a Lorentz function after subtracting a Shirley-type background, as shown in Figs. 4(d,e).
The peak position shifts from 42 meV to 22 meV with decreasing temperature and is saturated at 30 K.
The integrated intensity increases with decreasing temperature and is also saturated at 30 K.
According to photoelectron spectroscopic studies of bulk RE intermetallic compounds, such saturated temperature represents a coherent temperature ($T_{coh}$), at which the $c$-$f$ hybridization state is fully established, resulting in a HF state \cite{YbCuIn4, YbAgCu4}.
The above experimental fact strongly suggests that YbCu$_2$ is the 2D monoatomic-layered HF material with $T_{coh}$ = 30 K, which is more valence fluctuative than other low-dimensional HF such as CeIn$_3$/LaIn$_3$ superlattice ($T_{coh}$ = 1.6 K) \cite{CeIn3-2D}.

To investigate the momentum-dependent $c$-$f$ hybridization formation, we took the temperature-dependent peak position and intensity of the quasiparticle peak at three wavenumbers ($k_x$ = 0.5, 0.0, $-0.1$~\AA$^{-1}$) as shown in Figs. 4(d, e).
The change of the peak position at $k_x$ = 0.5 \AA$^{-1}$, which is the cut only the Yb$^{2+}$ 4$f_{7/2}$ state, almost follows the angle-integrated one.
This suggests that the angle-integrated spectrum mainly focuses on the high-density Yb$^{2+}$ 4$f$ state and the renormalization is effective for all 4$f$ states.
In contrast to the saturated feature in the angle-integrated spectrum at $T$ = 30 K, the KR peak positions at $k_x$ = 0.0 and $-0.1$~\AA $^{-1}$ are shifted toward the higher-binding energy side below $T_{coh}$ of YbCu$_2$ suggesting the hybridization gap enlargement, even though the peak shifts at $k_x$ = 0.0 and $-0.1$~\AA$^{-1}$ also follows the angle-integrated peak above $T_{coh}$.
Also, the peak intensity, which corresponds to the spectral weight transfer between Yb 4$f$ states and conduction bands, continued to increase below $T_{coh}$ as shown in Fig. 4(e).
To the best of our knowledge, similar behavior has not been reported yet except for the angle-integrated photoelectron spectroscopy of a Kondo semiconductor SmB$_6$ \cite{SmB6tshift, SmB6DMFT}, which originated from the transition from the metallic to the semiconducting state, but the magnitude of the peak shift observed here is much larger than that of SmB$_6$.
The peak shift in SmB$_6$ is immediately saturated just below $T_{\rm coh}$, which is not consistent with the behavior in YbCu$_2$, suggesting a different mechanism of the KR peak shift.
Both the peak shift and the developing intensity below $T_{coh}$ of YbCu$_2$ suggest that the 2D HF state still develops even below $T_{coh}$.
Further theoretical analysis such as the dynamical mean field theory about the development of the HF state in the 2D system would help the understand of the temperature-dependent behavior of the KR peak.

In conclusion, we report the electronic structures of a novel Yb-based monoatomic layered Kondo lattice YbCu$_2$ on Cu(111) by ARPES.
Our spectroscopic data provide direct evidence of the appearance of a purely 2D HF state with $T_{coh}$ = 30K, which is extremely higher than other 2D HF materials \cite{CeIn3-2D, CeRhInYbRhIn, He3}, in a monoatomic layer material for the first time.
Monoatomically layered YbCu$_2$ is the minimal material to realize low-dimensional HF containing RE elements and act as a building block to reveal novel electron-correlation-driven phenomena, for example, the proximity effect of layered material between 2D HFs and other many-body interactions such as superconductivity and magnetism.
Quantum fluctuations in 2D materials are much more sensitive to external fields \cite{CeRhInYbRhIn}.
The ground state of YbCu$_2$ would be tuned around QCP by external fields such as surface carrier doping by alkali metal adsorption and gate-tuning conventionally applied to other 2D materials in addition to a magnetic field and pressure, and is expected to explore novel quantum critical phenomena in 2D materials such as atomic-layer unconventional superconductivity.

\bibliographystyle{naturemag}
\bibliography{reference.bib}

\section*{Methods}
\subsection*{Sample preparation}
Cu(111) substrate was cleaned by Ar-ion sputtering with an acceleration energy of 0.5 keV and annealing at 800 K.
After several sputtering and annealing cycles, a sharp (1$\times1$) low-energy electron diffraction (LEED) pattern was confirmed as shown in Fig. 1(c).
Yb atoms were evaporated on the Cu(111) substrate at 600 K.
Because the crystallinity of the YbCu$_2$ layer was quite sensitive to the substrate condition such as cleanness of surface and growth temperature, we precisely monitored the sharpness of the diffraction from YbCu$_2$ and Moir\'{e} pattern, which is directly linked to the crystal quality of the YbCu$_2$/Cu(111), by the reflective high energy electron diffraction (RHEED) in the growth process.

\subsection*{Photoemission experiments}
ARPES and core-level photoelectron spectroscopy measurements were performed at BL-2A MUSASHI of the Photon Factory, and BL7U SAMRAI \cite{BL7U} of the UVSOR-III Synchrotron Facility.
The energy resolution and the energy position of the Fermi-level were calibrated by the Fermi-edge of polycrystalline Au films electrically contacted to the sample holder.
Energy resolutions for ARPES and core-level photoelectron spectroscopy were better than 20 meV and 100 meV, respectively.
In temperature-dependent measurements, the position of $E_{\rm F}$ and the instrumental resolution were accurately calibrated by measuring the Fermi edge of the Au thin film at all measurement temperatures.

\subsection*{Band calculations}
Band structures of freestanding YbCu$_2$ and YbCu$_2$/Cu(111) slab were calculated by using the WIEN2K code \cite{WIEN2k} including spin-orbit interaction within the generalized gradient approximation of the Perdew, Burke, and Ernzerhof exchange-correlation potential \cite{PBE}.
The in-plane lattice constant of the YbCu$_2$ was set to the experimentally obtained value (4.80 \AA) from LEED measurements.
The atomic structure of YbCu$_2$/Cu(111) was modeled by a symmetric slab of six layers of Cu with a surface covered with YbCu$_2$ layers.
No electron correlation was included, in the band calculations.
To obtain the overall trend of the electronic structure, such a condition that does not include the electron correlation would also be sufficient.
The calculated band structures are shown in Figs. S1 and S2.

\section*{Acknowledgement}
We acknowledge M. F. Lubis and K. Nishihara for their technical support during the experiments.
We would like to thank Professors Takahiro Ito and Hiroshi Watanabe for the helpful discussions.
The ARPES measurements were partially performed under UVSOR proposals 22IMS6861, 22IMS6848, and Photon Factory proposal 2022G513.
This work is supported by JSPS KAKENHI (Grants Nos.22K14605 and 20H04453).

\section*{Author Contributions}
T.N., H.S., and Y.C. conducted the ARPES experiments with assistance from R.Y., K.T. M.K., and H.K..
T.N. and Y.O. performed the DFT calculations.
T.N. and S.-i.K. wrote the text and were responsible for the overall direction of the research project.
All authors contributed to the scientific planning and discussions.

\section*{Data availability}
The datasets generated during and/or analyzed during the current study are available from the corresponding author upon reasonable request.

\begin{figure}[p]
\includegraphics[width=80mm]{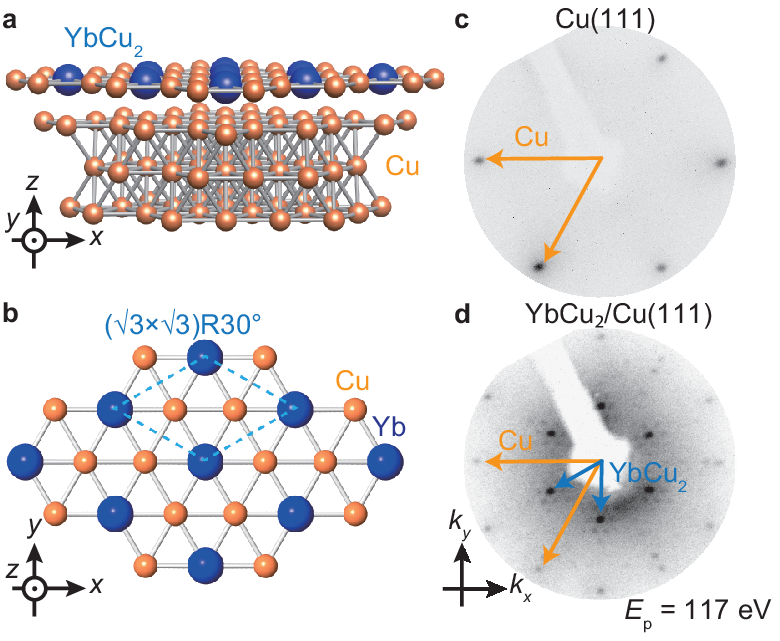}
\caption{\label{figure 1}
\textbf{Monoatomic-layer YbCu$_2$ on Cu(111) substrate.}
(a) A surface atomic structure of YbCu$_2$/Cu(111).
(b) Top view of monoatomic-layer YbCu$_2$.
The dashed line indicates the unit cell of YbCu$_2$.
(c) LEED pattern of Cu(111)-(1$\times$1) substrate.
(d) Same as (b) but for YbCu$_2$/Cu(111)-($\sqrt{3}\times\sqrt{3}$)R30$^{\circ}$.
Both LEED patterns were taken at the temperature of 70 K.
The primitive (1$\times$1) and ($\sqrt{3}\times\sqrt{3}$)R30$^{\circ}$ are indicated by orange and blue allows, respectively.
The distortions of the LEED image are due to the flat microchannel plate used for the LEED measurement.
The satellite spots around the integer spots represent the Moir\'{e} superstructure originating from a small lattice mismatch of YbCu$_2$ and Cu(111).
}
\end{figure}

\newpage
\begin{figure*}[p]
\includegraphics[width=150mm]{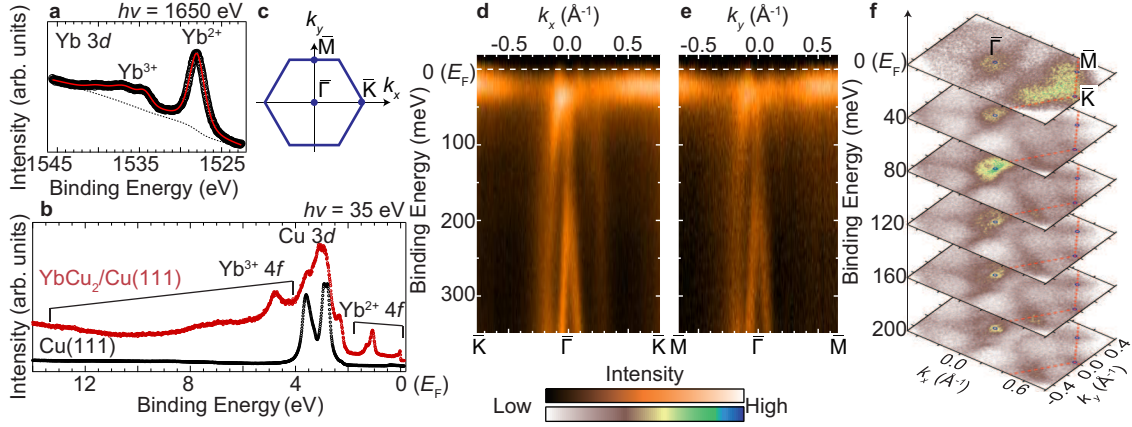}
\caption{\label{figure 2}
\textbf{Surface electronic structures of YbCu$_2$/Cu(111).}
(a) Yb 3$d$ core-level spectrum of YbCu$_2$/Cu(111) taken with 1650-eV photons at the temperature of 15 K.
Black circles and red lines represent the raw data and fitted curve, respectively.
The dotted line indicates the Shirley-type background.
(b) Angle-integrated valence-band spectra of Cu(111) (black) and YbCu$_2$/Cu(111) (red) taken with horizontally polarized 35-eV photons at 10 K.
(c) A hexagonal surface Brillouin zone of YbCu$_2$/Cu(111).
$k_x$ and $k_y$ are defined along $\bar{\Gamma}$--$\bar{\rm K}$ and $\bar{\Gamma}$--$\bar{\rm M}$ of YbCu$_2$.
(d,e) ARPES intensity plots along $\bar{\Gamma}$--$\bar{\rm K}$ and $\bar{\Gamma}$--$\bar{\rm M}$ taken with horizontally polarized 37-eV photons at 7 K.
ARPES intensities are divided by the Fermi--Dirac distribution function convolved with the instrumental resolution.
(g) Constant energy contours taken with the energy window of  $\pm$ 15 meV using 35-eV photons at 7 K.
The contour at the binding energy of 0 eV corresponds to an experimental Fermi surface.
}
\end{figure*}

\begin{figure}[p]
\includegraphics[width=80mm]{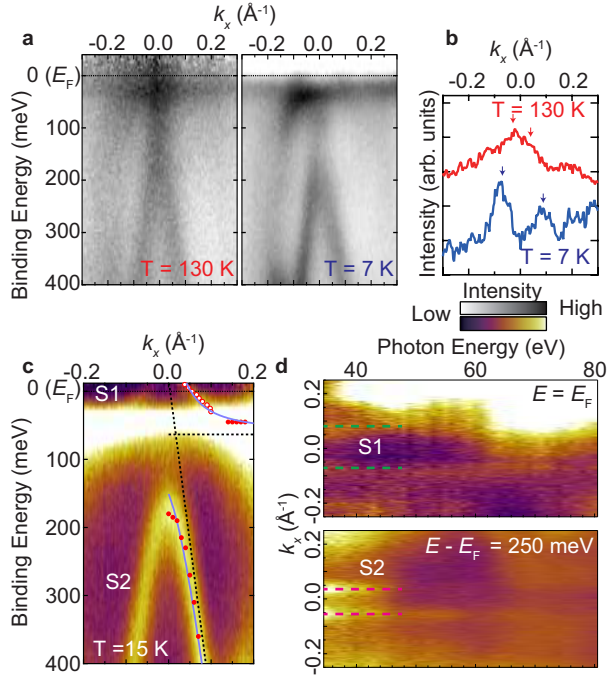}
\caption{\label{figure 3}
\textbf{Two-dimensional heavy fermion state in YbCu$_2$/Cu(111).}
(a) ARPES intensity plots along $\bar{\Gamma}$--$\bar{\rm K}$ at 130 and 7 K taken with horizontally polarized 37-eV photons.
ARPES intensities are divided by the Fermi--Dirac distribution function convolved with the instrumental resolution.
(b) Momentum distribution curves at $E_{\rm F}$ taken from (a) with the energy windows of $\pm$ 10 meV.
Arrows indicate the peak positions of the MDCs.
(c) Magnified ARPES image near the $\bar{\Gamma}$ point taken with circularly polarized 35-eV photons at 15 K.
ARPES intensities are divided by the Fermi--Dirac distribution function convolved with the instrumental resolution.
The filled and break lines indicate the simulated band dispersions ${E_k}^{\pm}$ with  $V_k$ = 120 meV and 0 meV by the PAM.
The open and filled circles indicate the peak positions from energy distribution curves (EDCs) and MDCs, respectively.
(d) Photon-energy dependence of MDCs at the normal emission at binding energies of 0 eV (upper panel) and 250 meV (lower panel) with the energy windows of $\pm$ 10 meV.
Dashed lines indicate the guide of the MDC peak position by eye.
}
\end{figure}

\begin{figure*}[p]
\includegraphics[width=150mm]{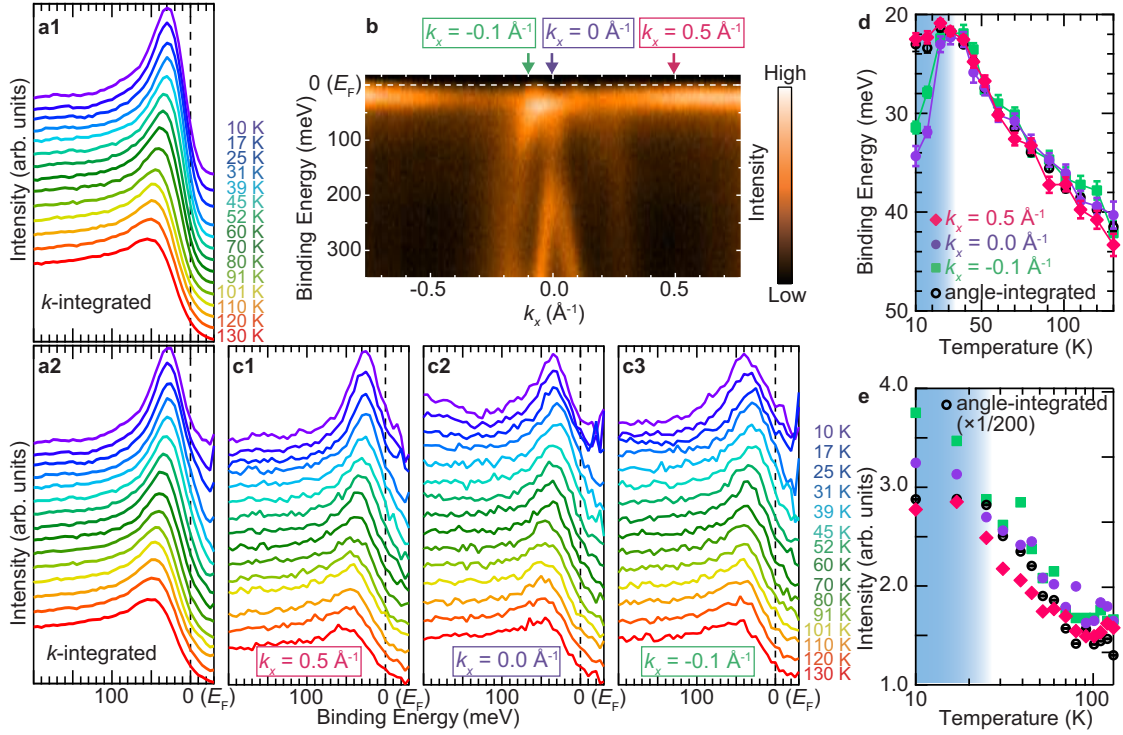}
\caption{\label{figure 4}
\textbf{Temperature- and momentum-dependent evolution of the Kondo-resonance peak.}
(a1) Angle-integrated photoelectron spectra near ${\rm E}_F$ as a function of temperature taken with horizontally polarized 35-eV photons.
(a2) same as (a1) but the intensity is normalized by the Fermi--Dirac distribution function convolved with the instrumental resolution.
(b) The same ARPES image as Fig. 2(d) indicates the $k_x$ positions where the temperature-dependent peak energy and intensity were measured.
(c) Angle-resolved photoelecron spectra near $E_{\rm F}$ as a function of the temperature.
The $k_x$ positions at 0.5, 0.0, and -0.1 \AA$^{-1}$ are representative of the local $f$ state only, the $\bar{\Gamma}$ point, and the crossing point of the $c$-$f$ hybridization, respectively.
(d) Momentum dependence of the energy position of the quasiparticle peak plotted on a linear scale of temperature.
(e) Momentum dependence of the intensity of the quasiparticle peak plotted on a logarithmic scale of temperature.
The shaded area indicates the HF state estimated by the saturated temperature in angle-integrated data, and the boundary between the shaded area and the white background region indicates the experimentally evaluated coherent temperature $T_{coh}$.
}
\end{figure*}

\newpage
\title{Supplementary Information for: Two-dimensional heavy fermion in a monoatomic-layer Kondo lattice YbCu$_2$}
\author{Takuto Nakamura}
\email{nakamura.takuto.fbs@osaka-u.ac.jp}
\affiliation{Graduate School of Frontier Biosciences, Osaka University, Suita 565-0871, Japan}
\affiliation{Department of Physics, Graduate School of Science, Osaka University, Toyonaka 560-0043, Japan}
\author{Hiroki Sugihara}
\affiliation{Department of Physics, Graduate School of Science, Osaka University, Toyonaka 560-0043, Japan}
\author{Yitong Chen}
\affiliation{Department of Physics, Graduate School of Science, Osaka University, Toyonaka 560-0043, Japan}
\author{Ryu Yukawa}
\affiliation{Graduate School of Engineering, Osaka University, Suita 565-0871, Japan}
\author{Yoshiyuki Ohtsubo}
\affiliation{National Institutes for Quantum Science and Technology, Sendai 980-8579, Japan}
\author{Kiyohisa Tanaka}
\affiliation{Institute for Molecular Science, Okazaki 444-8585, Japan}
\author{Miho Kitamura}
\affiliation{Photon Factory, Institute of Materials Structure Science, High Energy Accelerator Research Organization (KEK), 1-1 Oho, Tsukuba 305-0801, Japan}
\author{Hiroshi Kumigashira}
\affiliation{Institute of Multidisciplinary Research for Advanced Materials (IMRAM), Tohoku University, Sendai, 980–8577, Japan}
\author{Shin-ichi Kimura}
\email{kimura.shin-ichi.fbs@osaka-u.ac.jp}
\affiliation{Graduate School of Frontier Biosciences, Osaka University, Suita 565-0871, Japan}
\affiliation{Department of Physics, Graduate School of Science, Osaka University, Toyonaka 560-0043, Japan}
\affiliation{Institute for Molecular Science, Okazaki 444-8585, Japan}

\maketitle
\newpage
\renewcommand{\figurename}{Supplementary Figure}
\renewcommand{\thefigure}{S\arabic{figure}}
\setcounter{figure}{0}
\section*{Supplementary Note 1. Orbital contributions in the calculated band structure of YbCu$_2$}
In order to reveal the orbital character of the hybridized band, the band structure of the freestanding YbCu$_2$ is calculated as shown in Figure S1.
The flat bands at the binding energy of 0.2 and 1.5 eV are derived from Yb 4$f$ orbitals.
The hole band with the $\bar{\Gamma}$ point at its apex is mainly derived from Cu with a small contribution from the Yb 5$d$ state, which is consistent with the previous study \cite{GdAu22, YbAu2, ErCu2}.

\begin{figure*}[t]
\includegraphics[width=150mm]{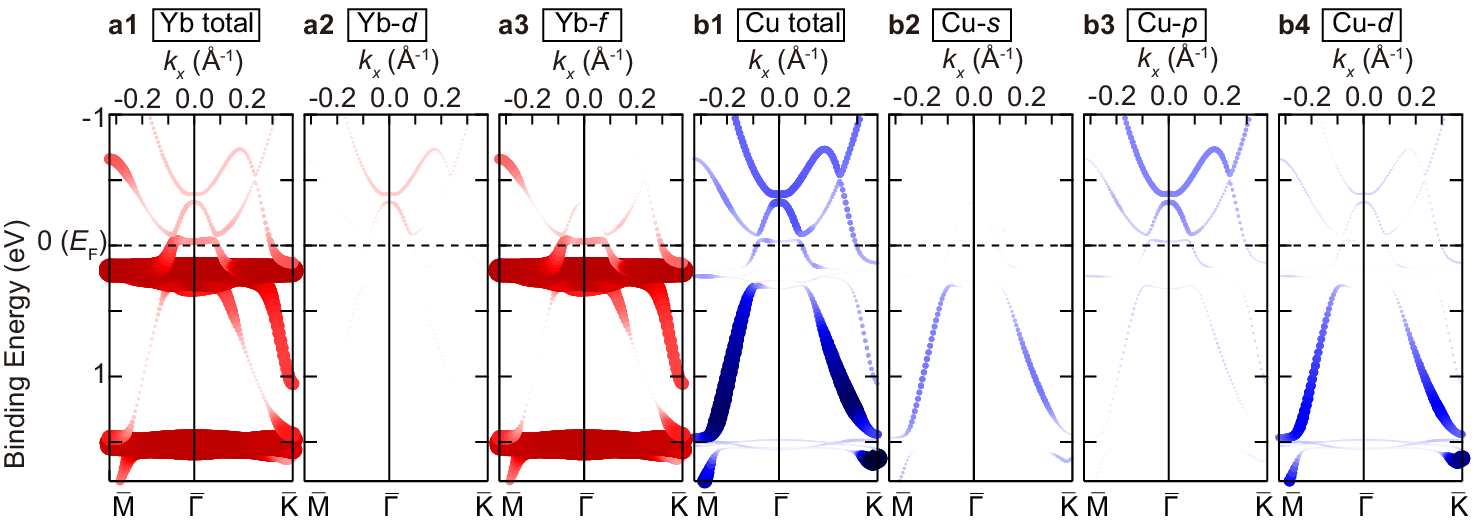}
\caption{
Orbitral contributions for the band structure of freestanding YbCu$_2$ by DFT calculation.
(a1-3) Band structure of freestanding YbCu$_2$, where the radii of the circles represent the projected contribution of (a1) all Yb, (a2) $f$- and (a3) $d$-orbitals.
(b1-4) same as (a) but for (b1) all Cu, (b2) $s$-, (b3) $p$- and (b4) $d$-orbitals.
}
\end{figure*}

The calculated band structures of the YbCu$_2$ on the Cu(111) substrate were shown in Fig. S2.
Overall shapes of the band structure are qualitatively consistent with the observed ARPES images.
The atomic orbitals in the innermost hole bands are mainly contributed from the Cu atoms in the YbCu$_2$ layer, indicating that these bands originate from the monolayer YbCu$_2$.
Note that other hole bands with the $\bar{\Gamma}$ point at its apex are derived from Cu atoms in the substrate.
These results indicate that the band structure of the monoatomic layered YbCu$_2$ still remains even including the Cu(111) substrate.

\begin{figure*}[t]
\includegraphics[width=120mm]{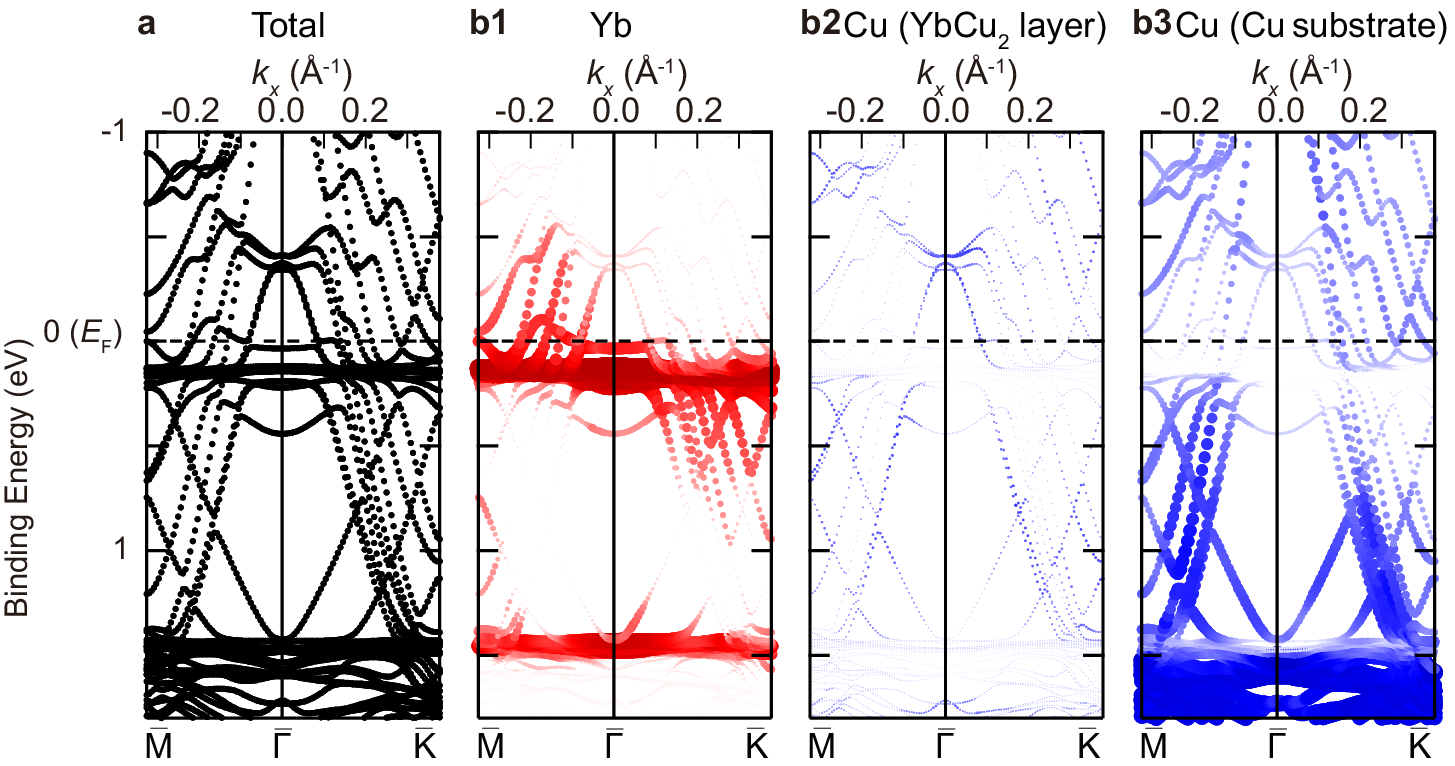}
\caption{
(a) Calculated band structure of the YbCu$_2$/Cu(111) slab.
(b)  same as (a) but the radii of the circles represent the projected contribution of (b1) Yb atom, (b2) Cu atom in the YbCu$_2$ layer, and (b3) Cu atom in the substrate.
}
\end{figure*}

\section*{Supplementary Note 2. Photon-energy dependence of MDCs}
Figure S3(a) shows the momentum distribution curves (MDCs) at $E_{\rm F}$ as a function of the excited photon energy.
The peaks at the $k_x$ = $\pm$ 0.08 \AA$^{-1}$, which correspond to the Fermi wavenumber of the hybridized conduction band S1 in Fig. 3(c) in the main text, are constant with changing the excited photon energy, indicating that the hybridized band is two dimensional.
Note that the peaks in MDCs excited at energies higher than 60 eV are not clearly visible due to the overlapping with other 3D hole bands as shown in Fig. S2(b3).
In order to confirm the contribution to the 2D HF state from the other hole band S3, which is the three-dimensional (3D) band described by the DFT calculation in Fig. S1, the photon-energy dependence of MDCs at the normal emission at binding energies of 100 meV is shown in Fig. S3(c).
The peaks at $k_x$ = $\pm$ 0.2 \AA$^{-1}$ in Fig. S3(b) depend on the excited photon energy suggesting the 3D character. This is consistent with the DFT calculations in Fig. S2(b3).
Note that the upper and lower branches of the $c$-$f$ hybridization band (S1 and S2 bands in Fig. S3(c)) have no photon-energy dependence suggesting the 2D character, which corresponds to Fig. 3(d) in the main text.

\begin{figure*}[h]
\includegraphics[width=150mm]{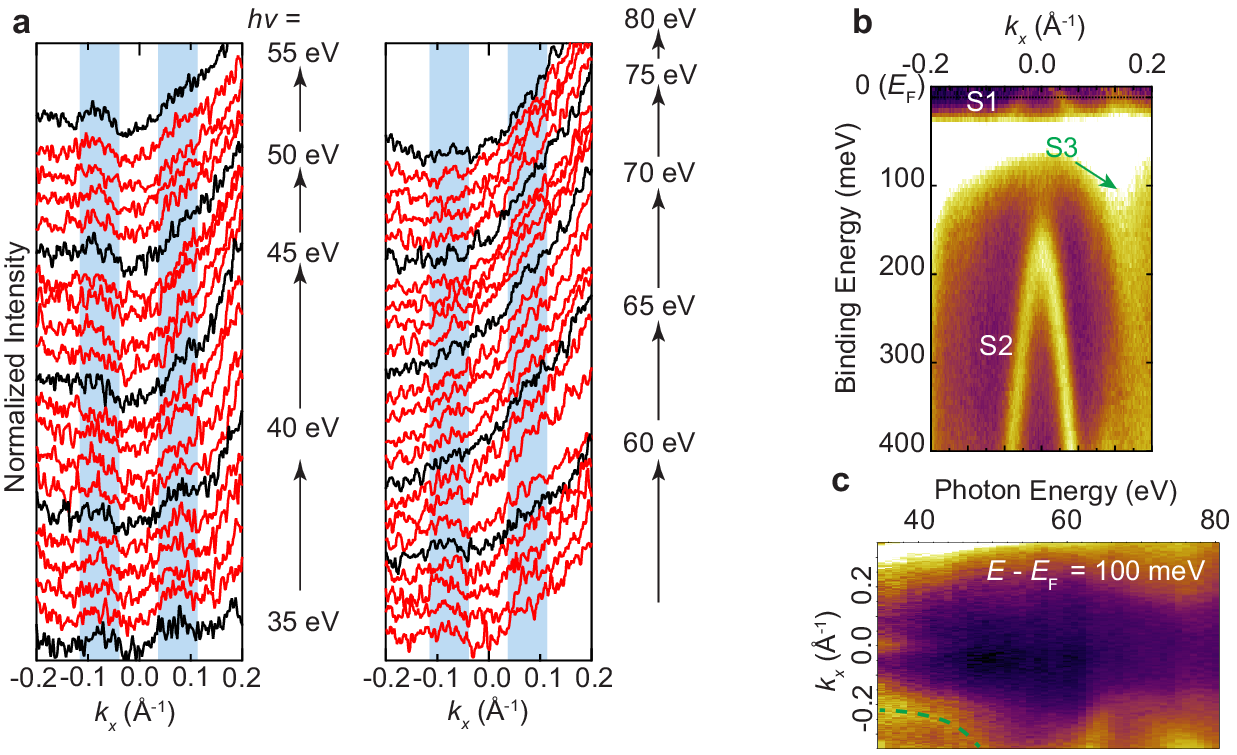}
\caption{\label{figure 2}
Momentum distribution curves (MDCs) at $E_{\rm F}$ with the energy windows of $\pm$ 10 meV at 15 K.
The incident photon energies ranging from 35 to 80 eV.
The intensities of all spectra are normalized by the average value of the peak at $k_x$ = $\pm$ 0.1 \AA$^{-1}$ to emphasize the peak energy position.
The shaded area highlights the peak position of the MDCs.
(b) Magnified ARPES image near the $\bar{\Gamma}$ point taken with circularly polarized 35-eV photons at 15 K.
ARPES intensities are divided by the Fermi-Dirac distribution function convolved with the instrumental resolution.
(c) Photon-energy dependence of MDCs at the normal emission at binding energies of 100 meV with the energy windows of $\pm$ 10 meV.
Dashed lines indicate the guide of the MDC peak position by eye.
}
\end{figure*}

\newpage
\section*{Supplementary Note 3. Fitting of Kondo-resonance peak in photoelectron spectra}
Figure S4 shows the temperature-dependent Kondo resonance peak with the fitting curves.
All spectra are fitted with a Lorentz function (purple line, an example for 130 K) after subtracting the Shirley-type background (black dotted line).
For the fitting of spectra at the low temperature at $k_x$ = 0.0 \AA$^{-1}$ in Fig. S4 (c), an additional Voight component was added to reproduce the peak at the binding energy of 180 meV (a down arrow), which originated from another 2D band S2 in Fig. 3 in the main text.

\begin{figure*}[h]
\includegraphics[width=150mm]{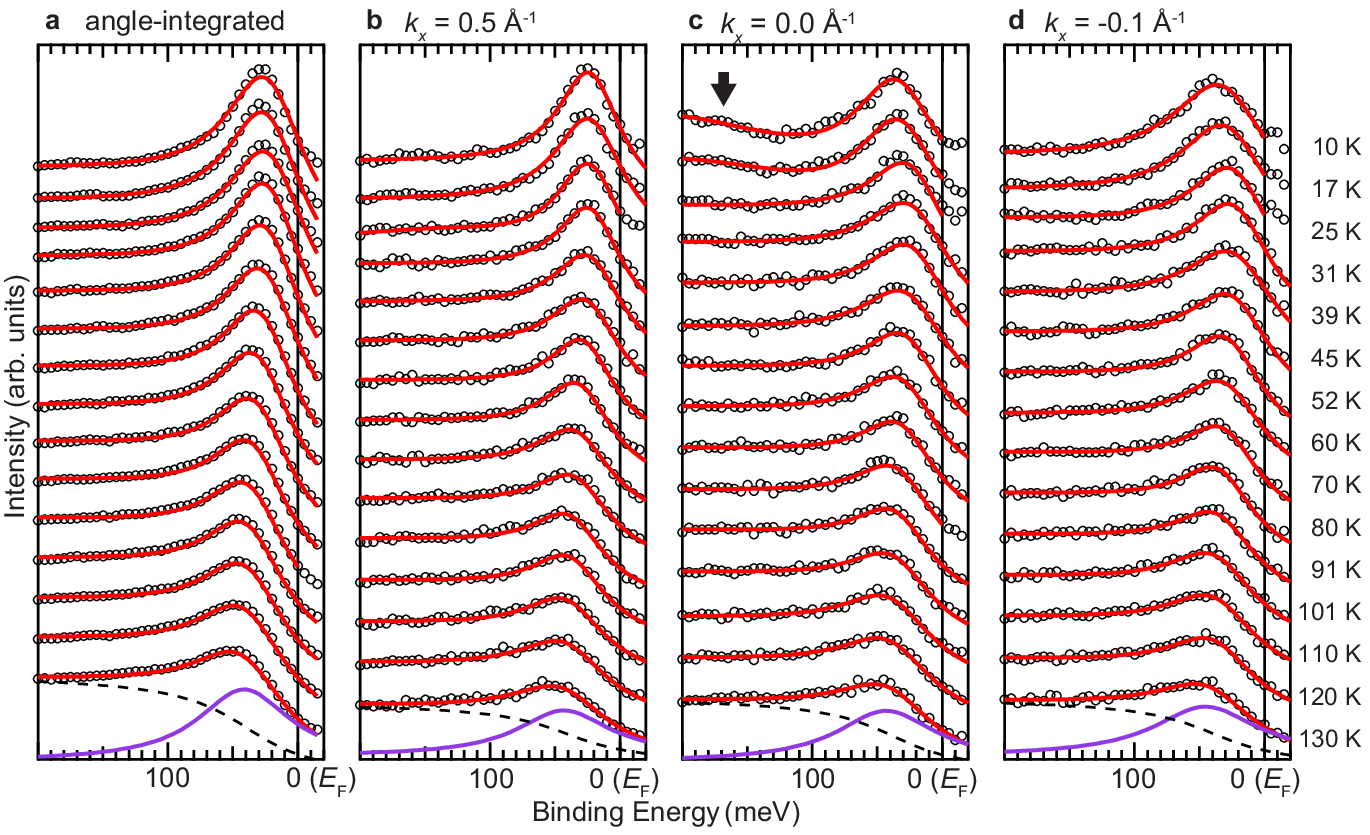}
\caption{\label{figure 4}
(a) Angle-integrated photoelectron spectra near $E_{\rm F}$ as a function of temperature taken with horizontally polarized 35-eV photons.
(b-d) Angle-resolved photoelecron spectra near $E_{\rm F}$ as a function of temperature.
The $k_x$ positions at (b) 0.5, (c) 0.0, and (d) -0.1 \AA$^{-1}$ correspond to the local $f$ state only, the $\bar{\Gamma}$ point, and the crossing point of the $c$-$f$ hybridization, respectively.
In all spectra, the intensity is normalized by the Fermi-Dirac distribution function convolved with the instrumental resolution.
Purple lines indicate Lorentz functions representing Kondo resonance peaks.
The dashed lines show Shirley-type backgrounds.
}
\end{figure*}
\newpage
\section*{Supplementary Note 4. Experimental geometry of angle-resolved photoelectron spectroscopy (ARPES)}
Figure S5 shows the geometry of the synchrotron-ARPES measurement in this study.
Photon-incident angle is 45 to $50^\circ$ respect to the normal of the photoelectron analyzer.
Both the photoelectron detection plane and the photon-incident plane are on the $xz$ plane.
It should be noted that the positive and negative photoelectron emission angles are non-equivalent geometric configurations in this study.

\begin{figure*}[h]
\includegraphics[width=150mm]{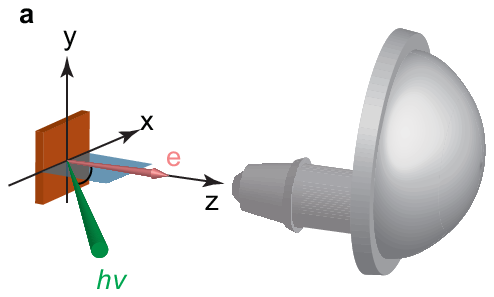}
\caption{\label{figure 2}
Schematic drawing of ARPES setup.
}
\end{figure*}

\section*{Supplementary Note 5. Evaluation of $c$-$f$ hybridization feature by periodic Anderson model}
To trace the $c$-$f$ hybridization parameters, the ARPES band dispersions were fitted with the Periodic Anderson Model (PAM) as shown in Fig. 3(c) in the main text.
From the fitting by Equation (1) in the main text, $\epsilon_f$ and $V_k$ are evaluated as 0.06 and 0.12 eV, respectively.
Here, the bare hole conduction band is assumed as a $k$-linear dispersion to explain the $c$-$f$ hybridization band dispersions.
From the PAM analysis, the Fermi velocity $v_{\rm F}$ and Fermi wavenumber $k_{\rm F}$ of the bare conduction band are evaluated as 4.77 eV ${\rm{\AA}}$ and 0.004 \AA$^{-1}$, respectively.
In order to confirm the validity of the bare bands by PAM analysis, an experimentally obtained ARPES image at room temperature is shown in Fig. S6 overlayed with assumed band dispersions with $V_k$ = 0 meV.
The shape of the assumed bare conduction band as well as the energy position of the 4$f$ band is in good agreement with the ARPES image.
From these parameters, the effective mass of bare conduction band $m_{b}$ becomes  5.65$\times$$10^{-33}$ kg.
On the other hand, by the fitting of the ARPES band dispersion near the $E_{\rm F}$ at 15 K, $v_{\rm F}$ and  $k_{\rm F}$ of the 2D HF band are 0.58 eV ${\rm{\AA}}$ and 0.055 \AA$^{-1}$, respectively, and the effective mass $m^{*}$ of the HF band is 6.56$\times$$10^{-31}$ kg.
Then,  the mass enhancement factor $m^{*}/m_{b}$ is evaluated as about 120.

\begin{figure*}[h]
\includegraphics[width=40mm]{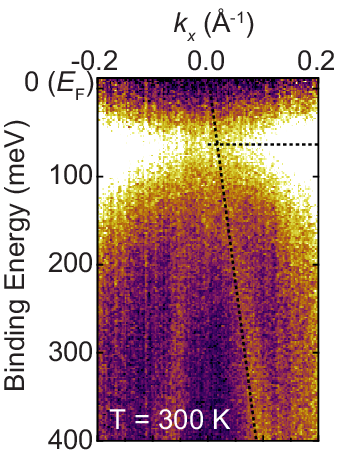}
\caption{\label{figure 6}
Magnified ARPES image near the $\bar{\Gamma}$ point taken with circularly polarized 35-eV photons at room temperature (300 K).
ARPES intensities are divided by the Fermi-Dirac distribution function convolved with the instrumental resolution.
The dashed lines indicate the assumed band dispersions ${E_k}^{\pm}$ with  $V_k$ =  0 meV in the PAM analysis.
}
\end{figure*}

\end{document}